\documentclass{webofc}
\usepackage[varg]{txfonts}   
\usepackage{comment}
\begin{document}
\title{Femtoscopy analysis in small systems at NA61/SHINE}

\author{\firstname{Barnab\'as} \lastname{P\'orfy}\inst{1,2}\fnsep\thanks{\email{barnabas.porfy@cern.ch, porfy.barnabas@wigner.hun-ren.hu}}
        for the NA61/SHINE Collaboration
}

\institute{Department of Atomic Physics, Faculty of Science, E\"otv\"os Lor\'and University, P\'azm\'any P\'eter s\'et\'any 1/A, H-1111 Budapest, Hungary
\and
           HUN-REN Wigner Research Centre for Physics, Konkoly-Thege Mikl\'os \'ut 29-33, H-1121 Budapest, Hungary
          }

\abstract{Recent measurements of femtoscopic correlations at NA61/SHINE unravel that the shape of the particle emitting source is not Gaussian. The measurements are based on L\'evy-stable symmetric sources, and we discuss the average pair transverse mass dependence of the source parameters. One of the parameters, the L\'evy exponent $\alpha$, is of particular importance. It describes the shape of the source, which, in the vicinity of the critical point of the QCD phase diagram, may be related to the critical exponent $\eta$. Its measurement hence may contribute to the search for and characterization of the critical point of the phase diagram. 
 
}
\maketitle

\section{Introduction}
\label{sec:intro}

In this proceedings, we report on our study using $^7$Be+$^9$Be and $^{40}$Ar+$^{45}$Sc collision systems with 150\textit{A} GeV/\textit{c} beam momentum ($\sqrt{s_{\textnormal{NN}}} = 16.82$ GeV), at $0$--$20$\% centrality in Be+Be and $0$--$10$\% centrality in Ar+Sc, collected by the NA61/SHINE experiment~\cite{Abgrall:2014xwa}. 

In our analysis we investigate the phase diagram of QCD using two-pion femtoscopic correlations with spherically symmetric L\'evy distributions, defined as:
\begin{equation}\label{eq:levydistr}
\mathcal{L}(\alpha,R,\textbf{r})=\frac{1}{(2\pi)^3} \int \textnormal{d}^3\vec{\zeta} e^{i\vec{\zeta} \textbf{r}} e^{-\frac{1}{2}|\vec{\zeta} R|^{\alpha}},
\end{equation}
where $R$ is the L\'evy scale parameter, $\alpha$ is the L\'evy stability index, $\textbf{r}$ is the vector of spatial coordinates and the vector $\zeta$ represents the integration variable.  The fundamental phenomenon utilized in femtoscopy is the connection between the spatial momentum correlations, $C(q)$ ($q$ is the relative momentum of the particle pair) and the spatial emission distribution $S(x)$. The latter describes the probability density of particle creation for a relative coordinate $x$. Then, the correlation function is expressed as $C(q) \cong 1 + | \tilde{S}(q) |^2$, where $\tilde{S}(q)$ is the Fourier transform of $S(x)$. Further details are available in Ref.~\cite{NA61SHINE:2023qzr}.

The L\'evy assumption allows us to describe the shape of the source with a more general approach. In the case of $\alpha=2$, one arrives to Gaussian distribution, while Cauchy distribution is obtained in the $\alpha = 1$ case.  Furthermore,  the L\'evy exponent is conjectured to be identical to the critical exponent $\eta$ related to spatial correlation~\cite{Csorgo:2005it}, as around the critical point (CP) spatial correlations will exhibit a power-law tail with an exponent of $(-1-\eta)$. Moreover,  L\'evy distribution also exhibits such a power-law tail $\sim r^{-1-\alpha}$, in case of $\alpha < 2$ (in three dimensions), where $r \equiv |\textbf{r}|$.  It has been suggested,  in Refs.~\cite{Halasz:1998qr,Stephanov:1998dy}, that the universality class of QCD is the same as that of the 3D Ising model in which case the expected value of $\eta$ around the CP is proposed to yield around $0.5$ or less~\cite{El-Showk:2014dwa, Rieger:1995aa}.  It is important to highlight that the appearance of L\'evy-shape can be attributed to several different factors besides critical phenomena, including jet fragmentation, anomalous diffusion, and others~\cite{Metzler:1999zz,Csorgo:2003uv, Csorgo:2004sr, Kincses:2022eqq, Korodi:2022ohn}. This motivates further measurements of $\alpha$ in different collision systems at various energies.
Then, the L\'evy exponent of the source distribution can be measured using two-particle femtoscopic correlation functions as $
C^0_2(q) = 1 + \lambda \cdot e^{-(qR)^\alpha},$
where $C^0_2(q)$ is the correlation function without final state phenomena.
The equation introduces an additional physical parameter, the correlation strength $\lambda$,  interpreted in the core-halo model in Refs.~\cite{Csorgo:1999sj,Csorgo:1995bi}.  In the limit as $q\rightarrow 0$, the correlation function converges to $1+\lambda$.  However, finite detector resolution forbids measuring $\lambda$ directly and an extrapolation from the region where two tracks are resolved becomes necessary.  The parameter can be expressed as $\lambda = \left(N_{\rm{core}}/(N_{\rm{core}}+N_{\rm{halo}})\right)^2$, with $N$ denoting their respective multiplicities, where the source function is divided into two components ($S = S_{\textnormal{core}} + S_{\textnormal{halo}}$).The core comprises pions created close to the center (primordial or strong decay pions), while the halo contains decay pions from long-lived resonances. 

The selection of like-charged pion pairs introduces a significant effect from the Coulomb repulsion. To account for this final state effect, the correlation function has to be modified to include the Coulomb correction (denoted by $K_{\textnormal{Coulomb}}$).  To this end,  the interpolation of Coulomb correction's functional form is used. The Coulomb correction, along with further final state corrections are described in greater detail in Refs.~\cite{Kincses:2019rug,Csanad:2019cns, Csanad:2019lkp, Kurgyis:2020vbz, NA61SHINE:2023qzr}. 

Subsequently,  the correlation function can be modified to incorporate both the Coulomb correction and the effect of the halo,with  the source radii of the core part remaining unaffected, as discussed in Ref.~\cite{Maj:2009ue}. To this end the Bowler-Sinyukov method~\cite{Sinyukov:1998fc,Bowler:1991vx} is utilized, modifying the fit ansatz as follows:
\begin{equation}
C_2(q) = N\cdot\left(1-\lambda+(1+e^{-|qR|^\alpha})\cdot\lambda\cdot K_{\textnormal{Coulomb}}(q)\right),\label{e:fittingformula}
\end{equation}
where $N$ is introduced as normalization parameter.  For the case of Ar+Sc, an additional parameter, $\varepsilon$,  is introduced, responsible for describing the linearity of the background,  taking the form of $(1+\varepsilon\cdot q)$. 

\section{Results}
\label{sec:results}

\begin{figure}
\centering
\includegraphics[width=0.49\textwidth]{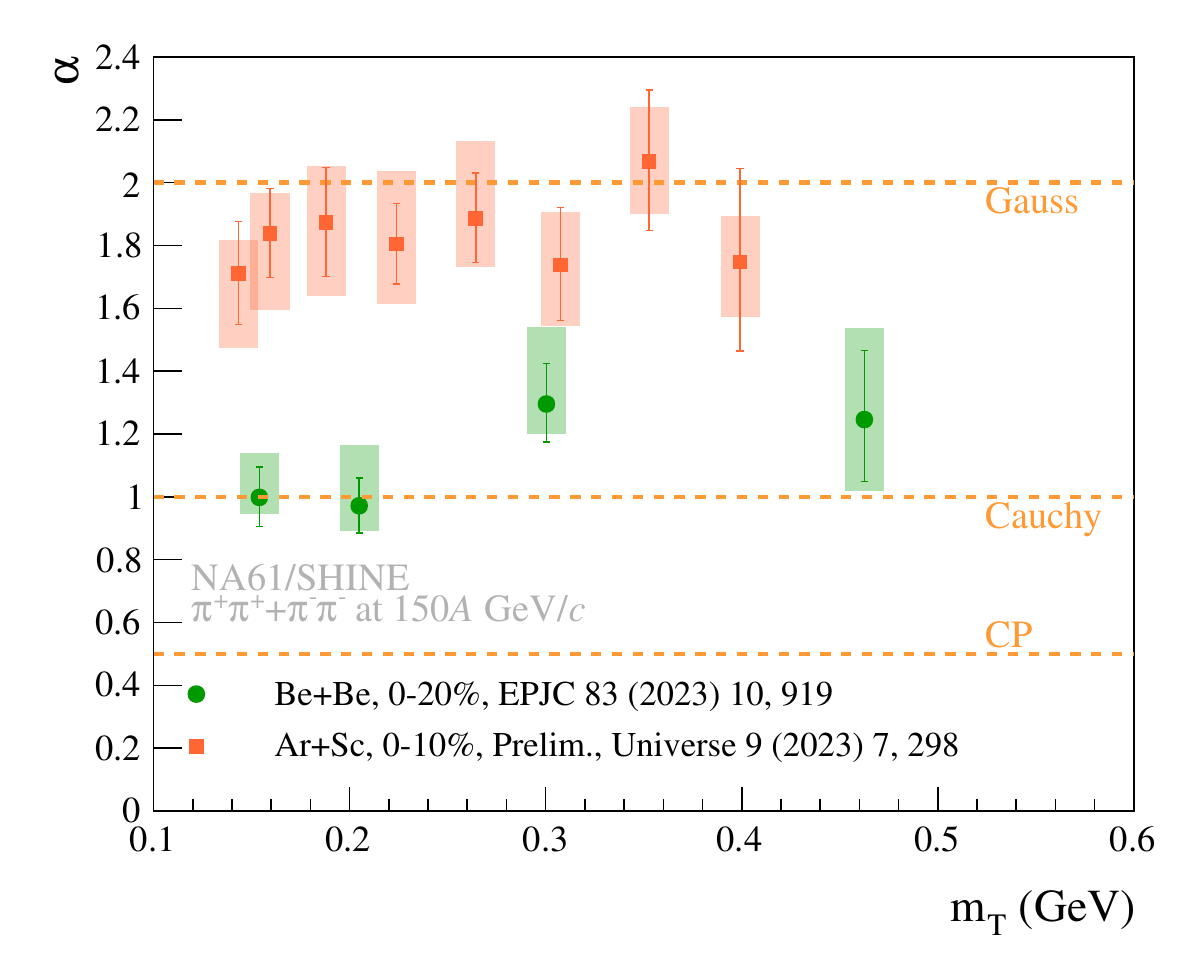}
\includegraphics[width=0.49\textwidth]{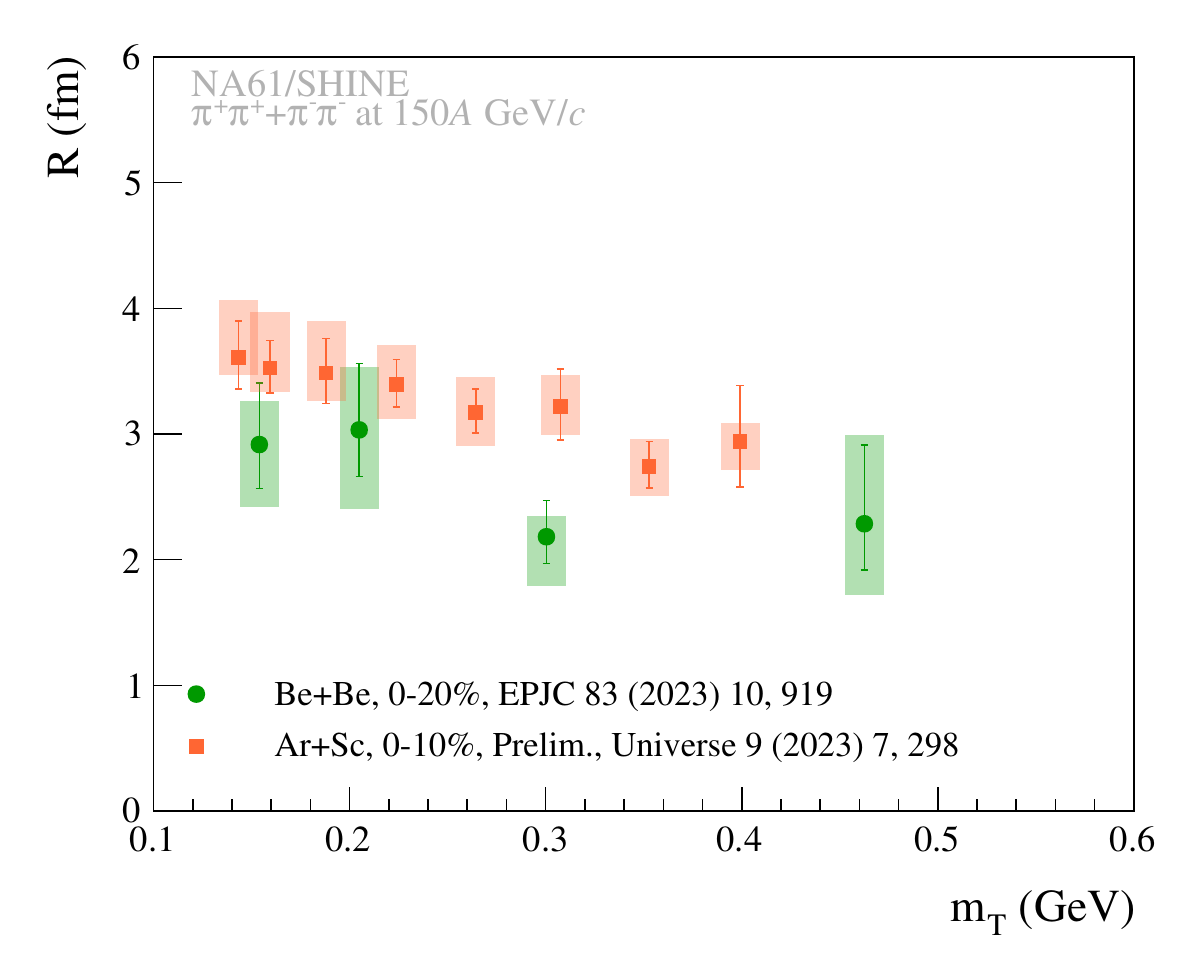}
\caption{The fit parameters, for $0$--$20$ \% central Be+Be at 150\textit{A} GeV/\textit{c} and $0$--$10$\% central Ar+Sc at 150\textit{A} GeV/\textit{c}, as a function of $m_{\textnormal{T}}$. Boxes denote systematic uncertainties, bars represent statistical uncertainties.}
\label{fig:results}
\end{figure}   

\textit{Femtoscopy}: We discuss the measurements of one dimensional two-pion femtoscopic correlation for identified pion pairs ($\pi^+\pi^+ + \pi^-\pi^-$) in Be+Be and in Ar+Sc collisions at 150\textit{A} GeV/\textit{c} with 0-20\% and 0--10\% centrality.  
The measured pion pairs were grouped into four (Be+Be) and eight (Ar+Sc) average transverse momentum bins ranging from 0 to 600 MeV/\textit{c} and from 0 to 450 MeV/\textit{c}, respectively. 

The $m_{\textnormal{T}} = \sqrt{m^2c^4+K_{\textnormal{T}}^2c^2}$ dependence of the physical parameters are shown on~Fig.\ref{fig:results}, where $K_{\textnormal{T}}$ is the average transverse momentum of the pair.  
The L\'evy stability exponent $\alpha$, can be used to extract the shape of the tail of the source. Our results yield values far from the value conjectured for the CP ($\alpha=0.5$), and that the source is not Gaussian ($\alpha=2$) nor Cauchy ($\alpha=1$) shaped. Together, they are suggesting that the measured correlation functions align with the assumption of a L\'evy source, indicating that it is more advantageous over the Gaussian assumption. 
The L\'evy scale parameter $R$ is related to the length of homogeneity~\cite{Sinyukov1995} of the pion emitting source. From simple hydrodynamical models, in Refs.~\cite{Csorgo:1995bi,Csanad:2009wc} one obtains a decreasing trend with transverse-mass for \textit{R}, and such slight decrease for higher $m_{\textnormal{T}}$ values may be observed, potentially caused by the transverse flow.  
We observe an $R(m_{\textnormal{T}})$ trend compatible with $R \sim 1/\sqrt{m_{\textnormal{T}}}$ prediction, which is particularly interesting as this type of dependence should rise in case of Gaussian sources~\cite{Sinyukov:1994vg}, observed also at RHIC~\cite{PHENIX:2017ino} and in simulations at RHIC and LHC energies~\cite{Kincses:2022eqq, Korodi:2022ohn}.  The intercept parameter $\lambda$ exhibits no dependence on $m_{\textnormal{T}}$. When compared to measurements from RHIC Au+Au collisions in Refs.~\cite{PHENIX:2017ino,Vertesi:2009wf, STAR:2009fks} and from SPS Pb+Pb interactions in Refs.~\cite{Beker:1994qv,NA49:2007fqa}, we can observe no visible ``holes'' at lower $m_{\textnormal{T}}$ values. This ``hole'' was interpreted in Ref.~\cite{PHENIX:2017ino} to be a sign of in-medium mass modification of $\eta'$. 

\textit{Intermittency}: Further analyses for the search of the critical point include the measurement of intermittent particle multiplicity fluctuations~\cite{Adhikary:2022sdh, Adhikary:2023rfj}. More precisely,  intermittency measurement aims to view the dependence of the second-order scaled factorial moments of proton multiplicity distributions on the number of subdivisions in transverse momentum space. The intermittency analysis uses statistically independent data sets for every subdivision in transverse and cumulative-transverse momentum variables~\cite{NA61SHINE:2023gez, NA61SHINE:2024ffp}. 
\begin{figure}[h!]
\centering
\includegraphics[width=0.49\textwidth]{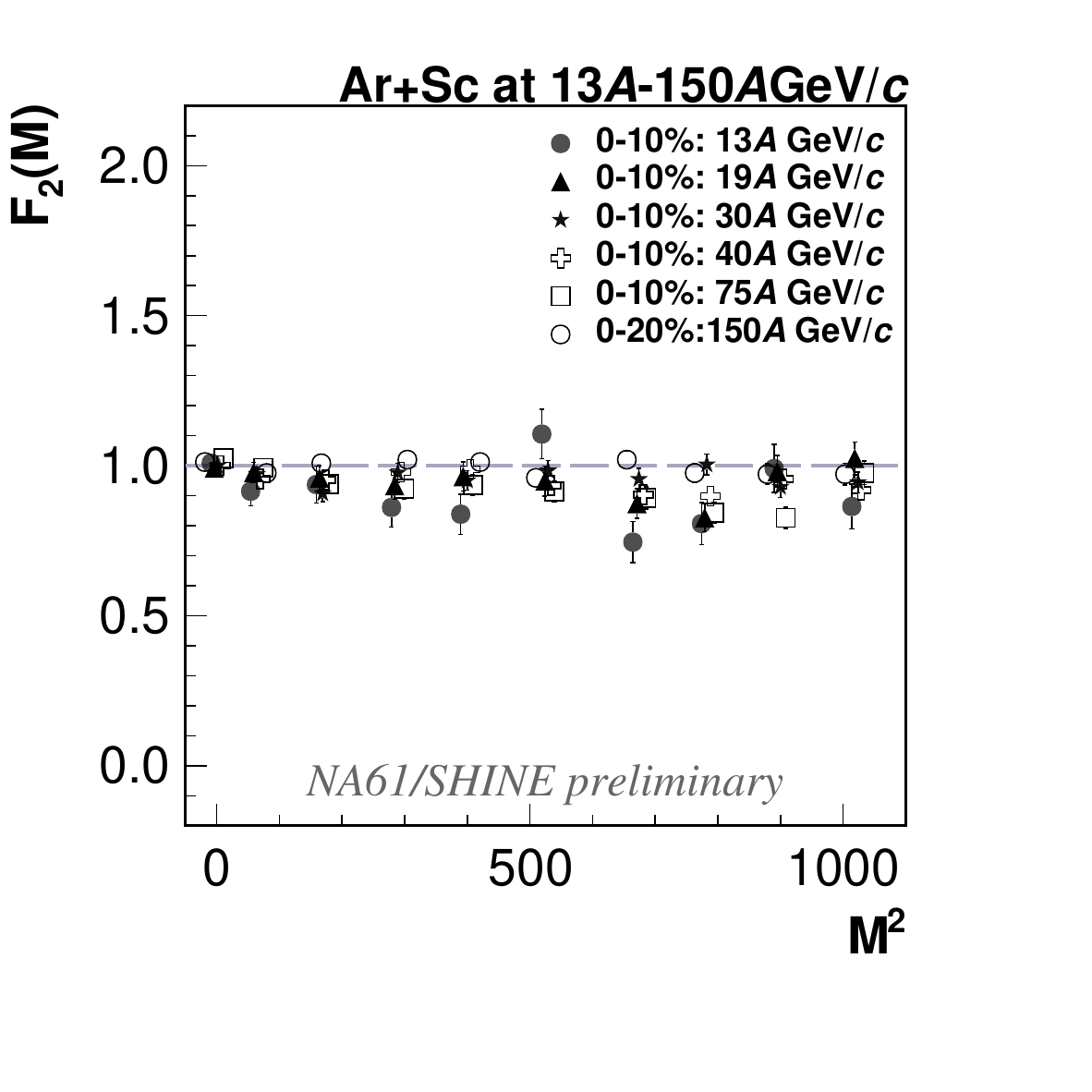}
\caption{Summary of the proton intermittency results from the NA61/SHINE Ar+Sc energy scan.}
\label{fig:int_results}
\end{figure}   
The scaled factorial moments $F_r(M)$ of order $r$ are defined by $F_r(M) = \frac{\left\langle \frac{1}{M^2}  \sum_{m=1}^{M^2} n_m(n_m - 1)... (n_m - r + 1) \right\rangle}{\left\langle \frac{1}{M^2} \sum_{m=1}^{M^2} n_m \right\rangle^r}$, where $M^D$ is the number of equally sized cells in D-dimensional space,  $n_m$ is the number of particles in m$^{th}$ bin and $\langle ... \rangle$ is the averaging over the events.  At the second order phase transition, the system is a simple fractal, and the factorial moment exhibits a power law dependence: $F_r(M) = F_r (\Delta) \cdot (M^D )^{\phi_q}$.
The results shown on Fig.~\ref{fig:int_results} do not indicate the searched intermittent pattern.

\textit{Acknowledgments}: The author acknowledges support of the DKOP-23 Doctoral Excellence Program of the Ministry for Culture and Innovation, and was furthermore supported by K-138136 and K-138152 grants of the National Research, Development and Innovation Fund. 

\bibliography{na61Preprint.bib}

\end{document}